\documentclass[10pt,epsfig]{article}
\usepackage{epsfig}
\newcommand{\be}{\begin{equation}}
\newcommand{\ba}{\begin{array}}
\newcommand{\bea}{\begin{eqnarray}}
\newcommand{\bfi}{\begin{figure}}
\newcommand{\ee}{\end{equation}}
\newcommand{\ea}{\end{array}}
\newcommand{\eea}{\end{eqnarray}}
\newcommand{\efi}{\end{figure}}

\newcommand{\lp}{\left(}
\newcommand{\rp}{\right)}
\newcommand{\ra}{\right\rangle}
\newcommand{\la}{\left\langle}
\begin{document}
%\baselikeskip 4ex

% Title page
\title{Intermittency in Turbulence: Multiplicative random process in space and time}
\author{Roberto Benzi$^{1,2}$, Luca Biferale$^{1,2}$  and  Federico Toschi$^{2,3}$\footnote{Corresponding author: Dr. Federico Toschi, IAC "Mauro Picone" (C.N.R.), Viale del Policlinico 137, 00161, Rome (ITALY), Phone: +39 06 88470262, Fax: +39 06 4404306.}\\
{\small $^1$ Dipartimento di Fisica, Universit\`a ``Tor Vergata'',}\\
{\small Via della Ricerca Scientifica 1, I-00133 Roma, Italy}\\
{\small $^2$ INFM, Sezione di Roma ``Tor Vergata'',}\\
{\small Via della Ricerca Scientifica 1, I-00133 Roma, Italy}\\
{\small $^3$ Istituto per le Applicazioni del Calcolo, CNR,}\\
{\small Viale del Policlinico 137, I-00161, Roma, Italy}}

%\pacs{47.27-i, 47.27.Nz, 47.27.Ak}
\maketitle
\pagestyle{myheadings}
\markboth{Intermittency in Turbulence: Multiplicative random process in space and time}{R. Benzi, L. Biferale and  F. Toschi\hfill Intermittency in Turbulence\dots}
\begin{abstract}
We present a simple stochastic algorithm for generating multiplicative processes 
with multiscaling both in space and in time. With this algorithm we are 
able to reproduce a synthetic signal with the same space and time 
correlation as the one coming from shell models for turbulence and the
one coming from a turbulent velocity field in a quasi-Lagrangian reference frame.
\vskip 0.2cm
\end{abstract}
\clearpage

\section{Introduction}
The multifractal language for turbulent flows   has been introduced about 20 years ago in order to 
describe the anomalous scaling properties of turbulence at large Reynolds numbers \cite{pf84,uriel}. Beside any particular interpretation, the multifractal formalism
exploit the scale invariance of the Navier-Stokes equation by taking into account fluctuations of the scaling exponents. To be more quantitative
let us consider the Navier-Stokes equations:
\be
\partial_t \vec u  +  \vec u \cdot \vec \nabla \vec u = - \frac{1}{\rho} \vec \nabla p + \nu \Delta \vec u
\ee
where $ \vec u $ is the velocity field describing a (homogeneous and isotropic) turbulent flow. For $\nu=0$ the Navier-Stokes equations
are invariant with respect to the scale transformation:
\be
\label{h}
r \rightarrow \lambda r \;\;\;\; u \rightarrow \lambda^h u \;\;\;\; t \rightarrow \lambda t^{1-h}
\ee
Then, following Kolmogorov, it is assumed that at large Reynolds numbers ($\nu \rightarrow 0$) the rate of energy dissipation is constant. As a consequence,
$h = 1/3$, if no fluctuation on $h$  are  present. 
The above reinterpretation of the Kolmogorov theory naturally opens  the way to describe intermittent fluctuations in turbulent flows.
Following the original idea by Parisi and Frisch, many  possible values of $h$ are allowed in turbulent flows. Each {\it fluctuation} $h$ 
at {\it scale} $r$ is weighted with a probability distribution $P_h(r) \sim r^{3-D(h)}$. \\
Since its first formulation, the multifractal model of intermittency have been applied to explain many statistical features of intermittency in a
unified approach.  The final goal of many theoretical investigation is to compute the function
$D(h)$ starting by the equation of motions. In some simple although highly non trivial case such a goal has been recently reached for the case of
 the Kraichnan model of a passive scalar \cite{rev}. \\
One of the key issue in the multifractal language of turbulence is to understand in a more constructive way what is a multifractal field and how the
fluctuations of $h$ are related to the dynamics of the system. In order to develop any systematic theory for computing $D(h)$ 
starting from the equation of motions, one has to handle a complex non linear problem: the way in which a perturbative scheme may be
developed strongly depend on a reasonable ansatz on the time-space properties of the probability distribution. It is therefore crucial to understand
how we can formulate the most general form of multifractal random field which is consistent with the time and space scaling properties of the
Navier-Stokes equations.    \\
One possible interpretation of the multifractal formalism is to observe that
for any $r < R$, the multifractal theory predicts:
\be
\delta u(r) = W(r,R) \delta u(R) \;\;\;\; \delta u(r) = u(x+r)-u(x)
\ee
Then, according to the scaling properties of $u$, the quantity $W(r,R)$ is a random quantity proportional to $(\frac{r}{R})^h$. It turns out that
for $r_1 < r_2 < r_3$ we have 
\be
W(r_1,r_3) = W(r_1,r_2) \cdot W(r_2,r_3)
\label{procmolt}
\ee
Equation (\ref{procmolt}) tells us that one possible interpretation of multifractal field is to assume that fluctuations at scale $r$ are described
by a random multiplicative process.  The random
multiplicative process is also somehow a simple way to mimic the energy cascade in turbulence. Actually, a general formulation of multifractal random
fields based on random multiplicative process was first presented in  \cite{physicad,pre} by using a wavelet decomposition of the field. \\
One obvious limitation of random multiplicative process is the absence of any time dynamics in the field, as one can immediately highlight by considering
space-time correlations. Space-time scaling is a crucial and delicate issue when considering multifractal fields for the Navier-Stokes equations \cite{proc1,bbct}. It is the aim of this paper to understand how one can exhibits a multifractal field whose space and time scaling
is consistent with the scaling constraints imposed by the Navier-Stokes equations. In Section 2 we introduce the technical problem shortly reviewed in this 
introduction by using a rather simplified language. In Section 3 we discuss several implications of the results obtained in Section 2, with a particular 
emphasis on the consequences for the fusion rules as introduced in \cite{eyink,procaccia_fr}. In Section 4 we outline our conclusions and we discuss future 
extensions of our research. 
\section{Multi-scale and multi-time stochastic signals}
\label{section2}
To simplify even further our argument we can concentrate an a typical fluctuation at a given scale, i.e. disregarding space position.
We introduce the scale hierarchy
$ l_n = l_0 \cdot \lambda^{-n} $, in terms of the scale separation $\lambda > 1$, and the velocity differences $w_n = v(x+l_n)-v(x)$.
We assume that the scaling properties of $w_n$ are consistent with the dimensional constraints imposes by the Navier-Stokes equations, i.e.
\be
\partial_t w_n \sim l_n^{-1} w_n^2
\label{nscale}
\ee
If $w_n$ shows multifractal scaling, we may write:
\be
\la w_n^p \ra \sim \la w_0\ra_0^p \lp {{l_n} \over {l_0}}\rp^{\zeta(p)}
\label{zetap}
\ee
where $\zeta(p)$ is a non linear function of $p$ and $\la\dots\ra_0$ is an
average over the large scale statistics. Following Parisi and Frisch, we know that
the multifractal scaling (\ref{zetap}) can be derived by assuming that $w_n \sim l_n^h$ with probability
$l_n^{3-D(h)}$, i.e.
\be
\la w_n^p\ra \sim \int dh \;l_n^{ph + 3 - D(h)} 
\ee
Indeed  by means of a saddle point evaluation of the previous integral, one obtain the explicit expression for $\zeta_p$ in terms of $D(h)$: 
\be
\zeta(p) = \inf_h \left[ph+3-D(h)\right]
\ee
Supposing one wants to keep into account also the time correlations, the constraint  (\ref{nscale}) implies that
\be
C_{p,q}(\tau) = \la w_n(t)^p w_n(0)^q\ra \sim \int dh l_n^{h(p+q)+3-D(h)} f_{p,q}(\tau/\tau_n) \;\;\; \tau_n = l_n/w_n
\label{time}
\ee 
where $\tau_n$ is a random time (the eddy turnover time) and the functions $f_{p,q}$ are dictated by the dynamic equations.
Expression (\ref{time}) has been introduced in  \cite{proc1} and analyzed in details in  \cite{bbct}.
We underline that, as a consequence of (\ref{time}), we can predict the 
scaling properties of quantities like $\frac{d^m C_{p,q}(\tau)}{d\tau^m}$.
We now want to understand how to define a random process satisfying both multiscaling in space (\ref{zetap}) and multiscaling in time (\ref{time}).
 In a more general way, we would like to exhibit random multifractal fields with prescribed dynamical scaling. 
It is known that the multifractal scaling (\ref{zetap}) can be observed for random multiplicative process. Let us introduce the (positive) random variable
$A_n$ and let us indicate with $P(A_n)$ the probability distribution of $A_n$. Then, by defining $w_n = \lp\prod_{i=1}^{n} A_i\rp w_0$ and by assuming that
 the random variables $A_i$ are independent, one obtains:
\be
\la w_n^p \ra \sim \int{\lp\prod_{i=1}^{n} A_i\rp^p \prod_{i=1}^{n} P(A_i) dA_i} = \la A^p\ra^n = l_n^{\zeta_0(p)} \;\;\; \zeta_0(p) = \log(\la A^p\ra)/\log{\lambda}
\label{rpm}
\ee
We want to generalize expression (\ref{rpm}) in order to satisfy the dynamical constrain (\ref{nscale}). At each scale $l_n$ we introduce the
random time $\tau_n = l_n/w_n$. 

The generation of our signal proceeds as follows, we extract $A_n$ 
with probability $P(A_n)$ and keep it constant for a time interval 
$[t,t+\tau_n]$. Thus, for each scale $l_n$, we introduce a time
dependent random process $A_n(t)$ which is piece-wise constant for a
 random time intervals $[t_n^{(k)},t_n^{(k)}+\tau_n]$, if $t_n^{(k)}$ 
is the time
of the $k$th jump at scale $n$. The corresponding 
 velocity field at scale $n$, in the time 
interval $t_n^{(k)} < t < t_n^{(k)}+\tau_n$,
is given by the simple multiplicative rule:
$$ w_n(t)= A_n(t)w_{n-1}(t_n^{(k)})$$
What is important to notice is that at each jumping time,
 $t_n^{(1)},t_n^{(2)},..,t_n^{(k)}..$
only the local velocity field is updated, i.e. information 
across different  scales propagates with a finite speed. 
 This is one possible and relatively simple way to take into
account the constraint (\ref{nscale}).

\begin{figure}
\begin{center}
\epsfig{file=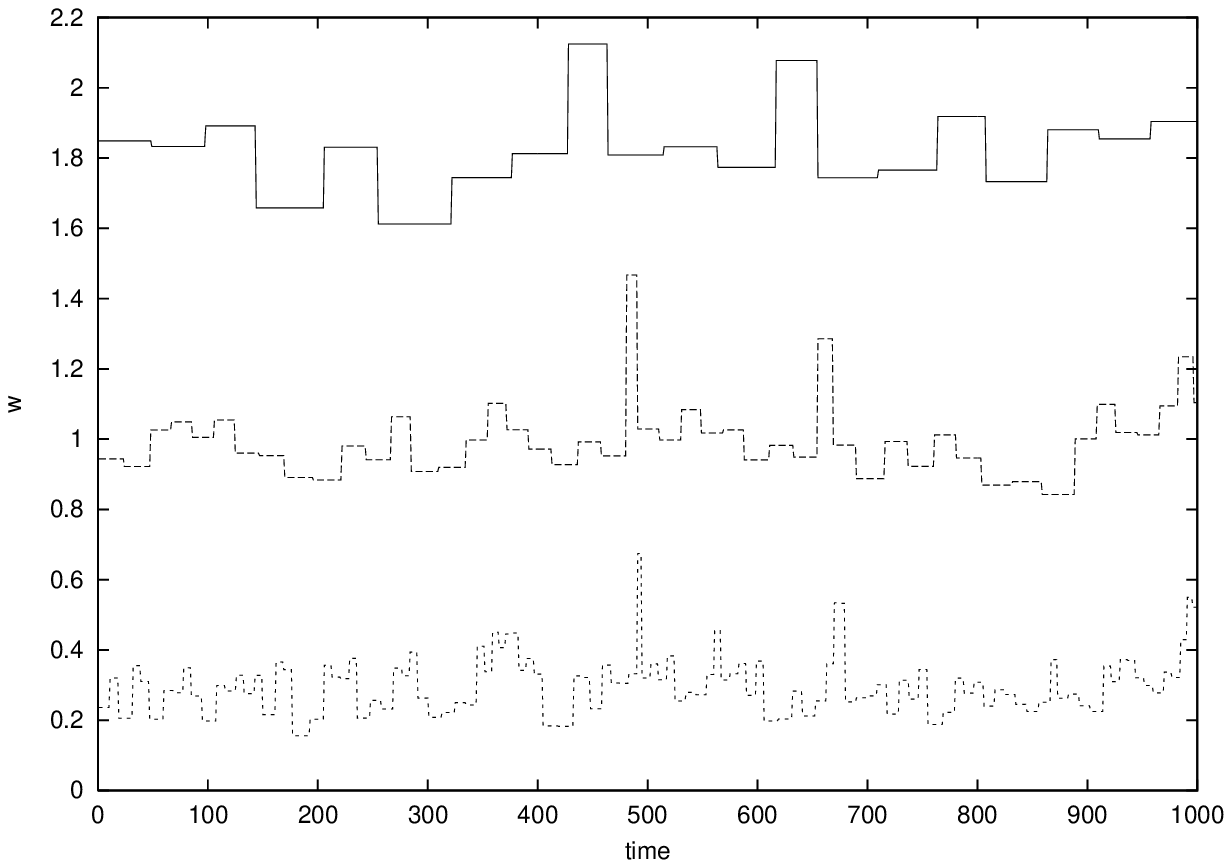,width=\hsize,keepaspectratio,clip}
\caption{Time behaviour of $w_n$ for $n=2,4,6$}
\label{fig0}
\end{center}
\end{figure}

To give a visual idea of how the algorithm works we show in Figure \ref{fig0} the time behaviour of the regeneration times for several scales. 
It is evident that there are short time-lag where the chain of multipliers is not given by an exact multiplicative process 
(this happens every time a small scale has 
to be regenerated but the ancestors are not yet dead).

\begin{figure}
\begin{center}
\epsfig{file=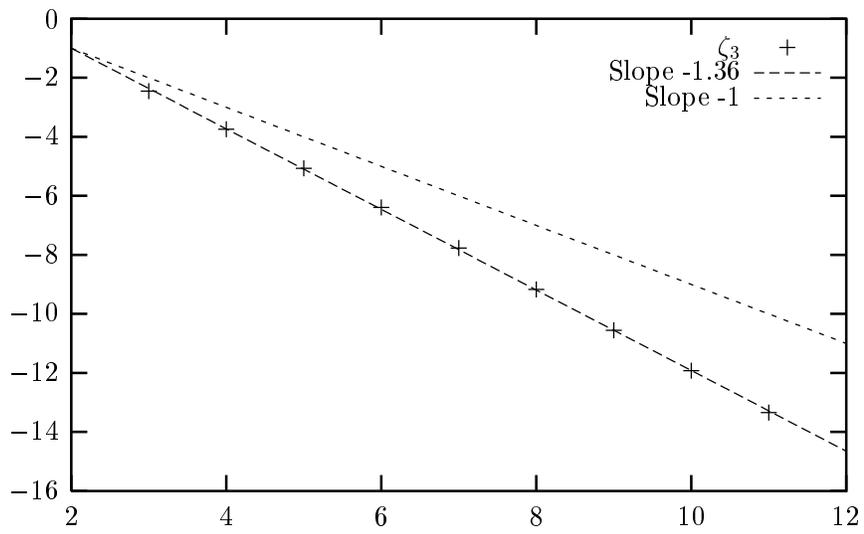,width=\hsize,keepaspectratio,clip}
\caption{Log-log plot of measured scaling of $S_3(n)$ vs. $n$ for a time dependent binomial random multiplicative process. 
The slope fitted slope $\zeta_3\simeq -1.36$ is ``renormalized'' with respect to the ``bare'' value $\zeta_3=1$.}
\label{fig1}
\end{center}
\end{figure}

In Figure \ref{fig1} we show the scaling behaviour of the third order structure functions
$S_3(l_n) = \la w_n^3 \ra $ obtained by a numerical simulation of a time dependent random multiplicative
 process with $P(A) = p_a\delta(A-A_a) + p_b\delta(A-A_b)$, 
where $A_a = 0.2$, $A_b=0.6$, $p_b= 1-p_a$ and $p_a$ has been chosen such that $\zeta(3) = 1$. Although $S_3$ shows a very well defined scaling,
 the value of $\zeta(3)$ is greater than what is predicted by (\ref{rpm}) (in Figure \ref{fig1} the slope $-1$ is shown for comparison). 
This effect shows that the ``real space'' scaling $\la w_n^p\ra$ is renormalized by the presence of the non trivial time dynamics of the multipliers.

In order to understand why (\ref{rpm}) cannot be used to predict the scaling exponents, 
let us understand which is the effect of the time dynamics for a given scale $l_n$ by assuming that at scale $l_{n-1}$ the 
variable $w_{n-1}$ is kept constant.
Let $T$ be the time used for time-averaged quantities and let  $N_a$ and $N_b$ be the number of events where the random variable $A$ is equal
 to $A_a$ and $A_b$ respectively. We next introduce the quantities
$\tau_a = l_n/(A_aw_{n-1})$ and $\tau_b = l_n/(A_bw_{n-1})$,
 the times associated
 to $A_a$ and $A_b$. By using our definition we can write: $N_a \tau_a + N_b \tau_b = T, \;\;\; N_a = p_a N, \;\;\; N_b = p_b N, \;\;\; N_a + N_b = N.$
It then follows:
\be
\la w_n^p\ra = \la w_{n-1}^p\ra \frac{1}{T}\int dt A(t)^p = \la w_{n-1}^p\ra \frac{1}{T} (\tau_a N_a A_a^p + \tau_b N_b A_b^p) 
\label{intermedia}
\ee
The above expression can be further simplified and we finally obtain:
\be
\la w_n^p\ra= 
\la w_{n-1}^p\ra \frac{p_a A_a^{p-1}+p_b A_b^{p-1}}{p_a A_a^{-1} + p_b A_b^{-1}}
\label{renorm}
\ee
The consequence of (\ref{renorm}) is that the scaling exponents $\zeta(p)$ are renormalized according to the expression:
\be
\zeta_R(p) = \zeta_0(p-1) -\zeta_0(-1)
\label{nzp}
\ee
where the number $\zeta_0(p)$ are the ``bare'' scaling exponents, i.e. those computed by using $P(A)$ according to (\ref{rpm}).

Expression (\ref{nzp}) have been obtained by using the simplified assumption $w_{n-1} = const$. In the general case, i.e. all variables $w_n$ are 
fluctuating, one needs to generalize the above discussion. A possible way is to write:
\be
\zeta_R(p) = \zeta_0(p-\alpha) -\zeta_0(-\alpha)
\label{newzp}
\ee
where the number $\alpha$ (not necessarily integer) depends on the details of $P(A)$.

\begin{figure}
\begin{center}
\epsfig{file=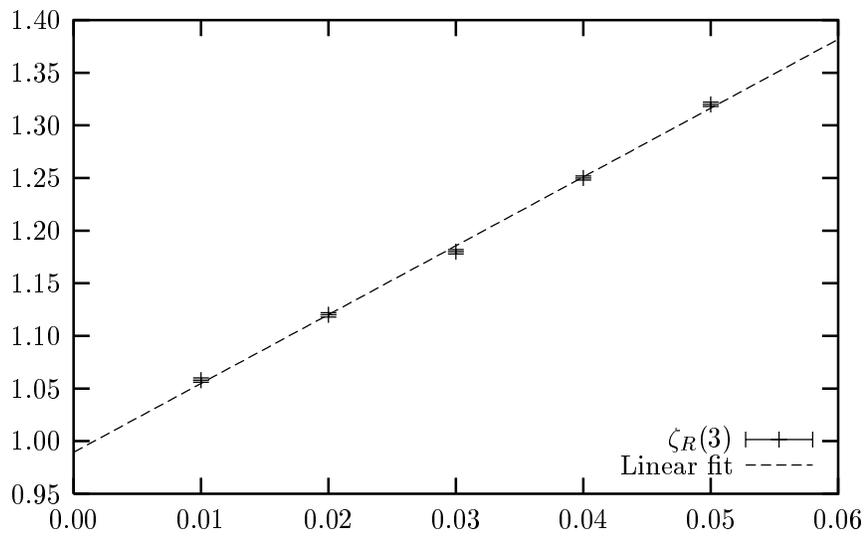,width=\hsize,keepaspectratio}
\caption{Behaviour of renormalized $\zeta_R(3)$ for log-normal distribution as a function of $\sigma^2$ (see text).}
\label{fig2a}
\end{center}
\end{figure}

\begin{figure}
\begin{center}
\epsfig{file=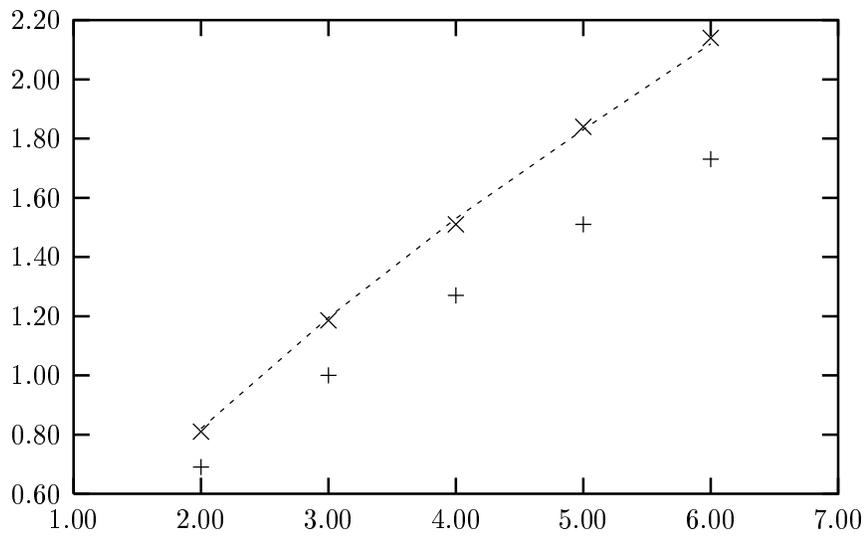,width=\hsize,keepaspectratio}
\caption{Behaviour of $\zeta_R(p)$ for $\sigma=0.03$ ($\times$) as compared to bare exponents $\zeta_0(p)$ ($+$). Dashed line is the prediction (\ref{lg1}).}
\label{fig2b}
\end{center}
\end{figure}

We have checked expression (\ref{newzp}) for a number of different choices of $P(A)$. Here we present the results for $P(A)$ being a log-normal
distribution, i.e. for 
\be
\zeta_0(p) = p h_0 - \frac{1}{2} \sigma^2 p^2
\label{lg0}
\ee
By using (\ref{newzp}) we obtain:
\be
\zeta_R(p) = p(h_0+\alpha\sigma^2) -  \frac{1}{2} \sigma^2 p^2
\label{lg1}
\ee
In Figure (\ref{fig2a}) we show $\zeta_R(3)$ obtained by a set of numerical simulations 
for different values of $\sigma$ and $h_0$ chosen in such a way that $\zeta_0(3) = 1$. In this case (\ref{lg1}) can be written as
$\zeta_R(3) = 1+3\alpha\sigma^2$. 
As one can easily observe, the prediction of (\ref{lg1}) is verified with very good accuracy with a value of $\alpha$ close to $2$. 
In Figure (\ref{fig2b}) we show the value of $\zeta_R(p)$ for $p = 2,..,6$ as obtained by direct numerical simulation for $\sigma = 0.03$. 
The dashed line represent the estimate (\ref{lg1}) with $\alpha = 2$: a very good agreement is observed.

\begin{figure}
\begin{center}
\epsfig{file=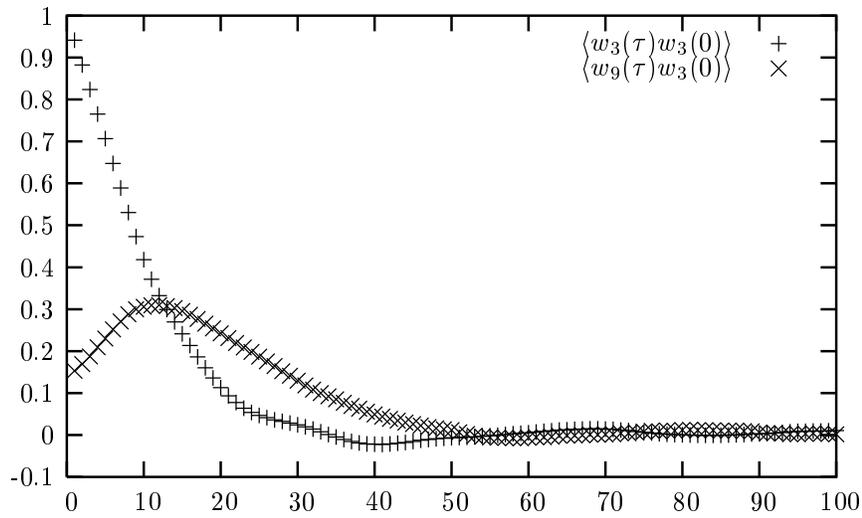,width=\hsize,keepaspectratio}
\caption{Time correlations at the same scale, $\la w_3(\tau) w_3(0)\ra$, and at different scale,$\la w_9(\tau) w_3(0)\ra$, as a function of the time lag $\tau$.}
\label{fig6a2}
\end{center}
\end{figure}

The above discussion can be generalized for multifractal fields and for any particular choice of $P(A)$. In all cases, a re-normalization of the
scaling exponents, as predicted by (\ref{lg1}), should be expected. At the same time, a non trivial time correlation is introduced for the variables
$w_n$. In Figure (\ref{fig6a2}) we show $\la w_n(\tau)w_n(0)\ra$ and $\la w_{n+6}(\tau)w_n(0)\ra$ for $n=3$, obtained by a numerical simulation of 
the time dependent random multiplicative process with a log-normal distribution with parameters $\sigma= 0.03$, $\zeta_R(3) = 1$. As expected, the correlation  
$\la w_{n+6}(\tau)w_n(0)\ra$
increases for small $\tau$ and then goes to $0$. The pick at a time lag larger than zero is due to the presence of a non trivial time dynamics, i.e.
multipliers at different scales need some time to realize that their ancestor have changed their status.  This is meant to mimic the 
non-trivial time dynamics of the turbulent energy transfer. This behaviour is also observed in the numerical simulation of deterministic shell models as reported in \cite{bbct}.

\begin{figure}
\begin{center}
\epsfig{file= 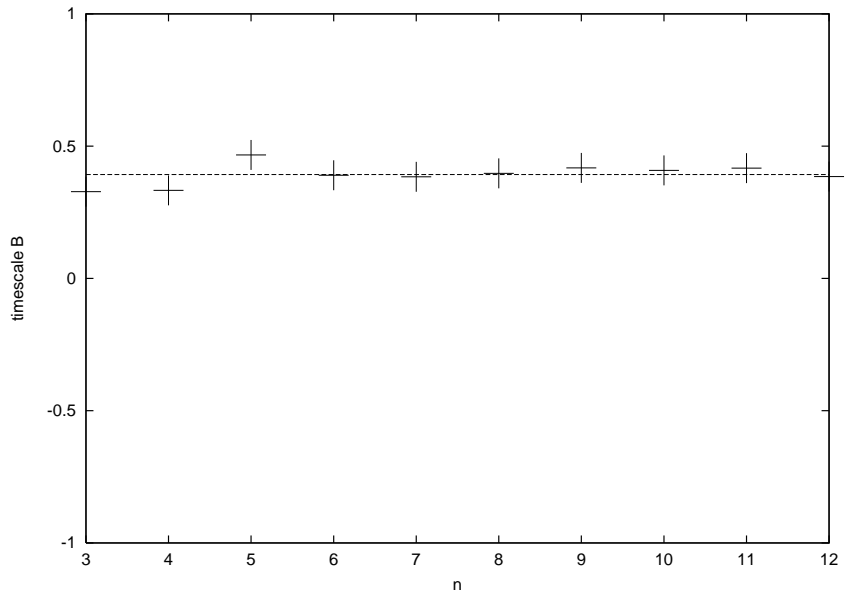,width=\hsize,keepaspectratio}
\caption{The time scale $B$, computed by the expression $\la w_n(t+\tau)w_n(t) \ra \sim \exp{(-B|\tau|)}$, for different value of $n$. The scaling constrain 
(\ref{nscale}) should correspond to $B\sim const$ as observed.}
\label{B}
\end{center}
\end{figure}

Finally let us check whether the quantities $w_n$ satisfies the scaling constrain imposed by (\ref{nscale}). We first observe that the correlation function
$\la w_n(t+\tau) w_n(t) \ra$ goes as $\exp{(-B|\tau|)}$. This is due to the fact $w_n$ as a function of time is not differentiable. In order to check whether 
(\ref{nscale}) is satisfied, we observe that $B \sim k_n w_n^3 \sim const$ if $\zeta_R(3) = 1$. In Figure (\ref{B}) we plot the timescale $B$ as a function
of $n$ obtained by a time dependent random multiplicative process, using a log normal distribution with $\sigma = 0.03$, $\zeta_0(3) = 0.83$ and $\zeta_R(3)=1$. As one can see, $B$ is fairly constant in the inertial range.

\section{Numerical results}
\begin{figure}
\begin{center}
\epsfig{file= 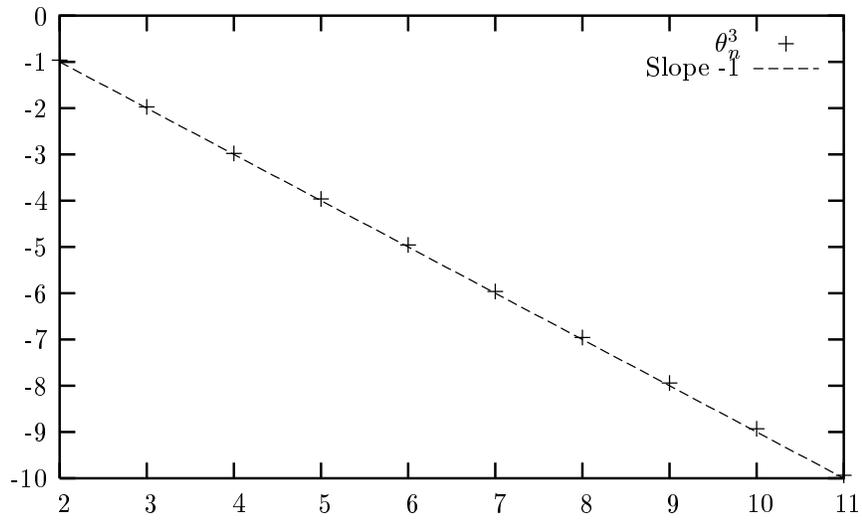,width=\hsize,keepaspectratio,clip}
\caption{Log-log plot of $\la \theta_n^3\ra$ vs $n$, for the passive scalar as compared to the theoretical prediction with slope $-1$.}
\label{fig4}
\end{center}
\end{figure}

The re-normalization effects discussed in the previous Section can be further investigated by considering the case of a passive scalar.
In this case the constraint (\ref{nscale}) should be written as:
\be
\partial \theta_n \sim \theta_n \frac{w_n}{l_n}
\label{nscalep}
\ee
where $\theta_n = \theta(x+l_n) - \theta(x)$ in analogy with the definition of $w_n$. We assume that 
$\la \theta_n^p\ra \sim l_n^{H(p)}$.
Let us assume that a suitable representation of $\theta_n$ is
given by a time dependent random multiplicative process as described in Section \ref{section2}. In this case, however, the random time $\tau_n$ is not correlated
with the value of $\theta_n$ or the related random multiplicative variables. Therefore, we should not expect any re-normalization for the scaling exponents
$H(p)$. This is indeed the case as shown in  
Figure (\ref{fig4}), where  we plot the scaling of $\la \theta_n^3\ra$ for a log-normal random multiplicative process with $\sigma = 0.03$ and $h_0$ such that
$H(3) = 1$. The updating times have been chosen with an independent random distribution mimicking the velocity fluctuations, $\tau_n \sim l_n/w_n$
 The dashed line corresponds to a slope $-1$.

Our definition of time dependent random multiplicative process can be very useful in investigating the behaviour of the so called fusion rules for the
quantities like $\la w_{n+m}^pw_n^q \ra$. Following  \cite{eyink,procaccia_fr,prl}, we can write:
\be
\la w_{n+m}^pw_n^q \ra = C_{p,q}(m) \frac{\la w_{m+n}^p\ra}{\la w_n^p\ra} \la w_n^{p+q}\ra
\label{fusion}
\ee
where $C_{p,q}$ is a constant for large $m$. Actually a direct measurements of $\la w_{n+m}^p w_n^q \ra$
in turbulent flows at high Reynolds numbers and in direct numerical simulations, confirm the validity of (\ref{fusion}) with  $C_{p,q}<1$ for 
large $m$ \cite{prl}. 
It is interesting to observe that our time dependent random multiplicative process satisfies fusion rules with  $C_{p,q}<1$. It is indeed possible
to show this result by the following argument.
Let us consider two scales $l_n$ and $l_{n+m}$. For fixed time, 
the quantity $w_n$ and
$w_{n+m}$ are not necessary product of the same random variables. They
will  {\it feel}  the same chain of multipliers only for   some larger 
scale $l_{n'}$ with $n'<n$, while for smaller scales both chain of multipliers
leading to the instantaneous value of $w_n$ and $w_{n+m}$ may have changed.
 Thus we should have:
\be
\la w_n^qw_{n+m}^p\ra = \la \prod_{k=1}^{n'} A_k^{q+p}\ra \la\prod_{k=n'}^{n}A_k^q\ra \la \prod_{k=n'}^{n+m}A_k^p\ra
\ee
where $A_k$ is the random multiplier acting between scale $k-1$ and scale $k$, i.e. $w_{k+1} = A_k w_k$. The above
expression gives the following result for the compensated fusion rules:
\be
{{\la w_n^qw_{n+m}^p\ra \la w_n^p\ra} \over {\la w_{n+m}^p\ra \la w_n^{p+q}\ra}}  \sim 
{\la A^{q+p}\ra^{n'} \la A^q\ra^{n-n'}\la A^p\ra^{m+n-n'} \la A^p\ra^n \over \la A^p\ra^{n+m} \la A^{p+q}\ra^n} 
\ee
that is:
\be
{{\la w_n^q w_{n+m}^p\ra \la w_n^p\ra} \over {\la w_{n+m}^p\ra \la w_n^{p+q}\ra}} = \left( {{\la A^{q+p}\ra} \over {\la A^q\ra}{\la A^p\ra}}\right)^{n'-n} \le 1 
\label{cpq}
\ee
Let us notice that for $n'=n$, which corresponds to random multiplicative process without time dynamics, the r.h.s of the above expression
is just $1$. The above equation should be considered as a qualitative prediction. In general we expect $n'$ to be a function
of both $p$ and $q$.

\begin{figure}
\begin{center}
\epsfig{file=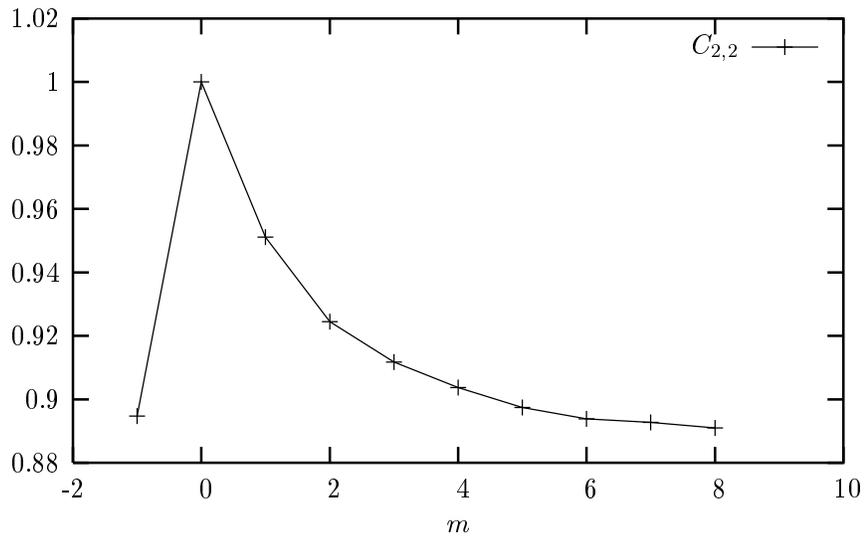,width=\hsize,keepaspectratio,clip}
\caption{Behaviour of $C_{2,2}(m) = \frac{\la w_{3+m}^2w_3^2 \ra \la w_3^2 \ra}{ \la w_{3+m}^2 \ra \la w_3^4\ra}$, defined by using (\ref{fusion}) for $n=3$, as a function of $m=-1,0,1,...$, for the log-normal probability. 
For $m=0$ $C_{2,2}(0) = 1$ while for large and positive $m$ $C_{2,2}$ reaches a plateau smaller than $1$.} 
\label{fig6a1}
\end{center}
\end{figure}

We have checked our prediction by several simulations for different choices of $P(A)$. In Figure (\ref{fig6a1}) we show the quantity
$C_{2,2}(m)$ defined as in (\ref{fusion}) for $n=3$ and for 
$m=-1,..11$. As expected $C_{2,2}(m)=1$ for $m=0$ and for $m>0$
 is a slowing decaying 
function of $m$ which reaches a plateau only for very large $m$.

\begin{figure}
\begin{center}
\epsfig{file=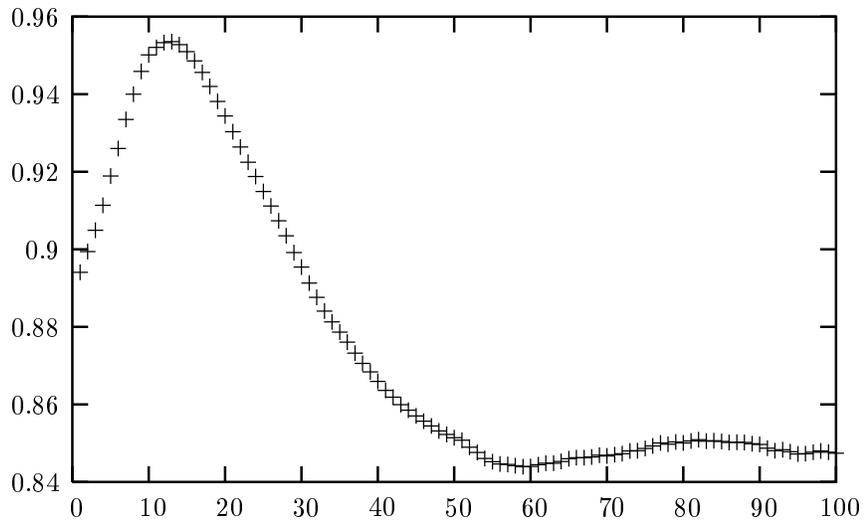,width=\hsize,keepaspectratio,clip}
\caption{Behaviour of $C_{9,3}(t)$ as a function of time.}
\label{fig6a3}
\end{center}
\end{figure}

As a consequence of our analysis, we can also predict that the quantity $C_{p,q}(t)$, defined through:
$$
\la w_{n+m}(t)^pw_n(0)^q\ra = C_{p,q}(t) \frac{\la w_{m+n}^p\ra}{\la w_n^p\ra} \la w_n^{p+q}\ra \;\;\; m \gg 1
$$
should increase with $t$. This is indeed the case as one can observe in Figure (\ref{fig6a3}).

\begin{figure}
\begin{center}
\epsfig{file= 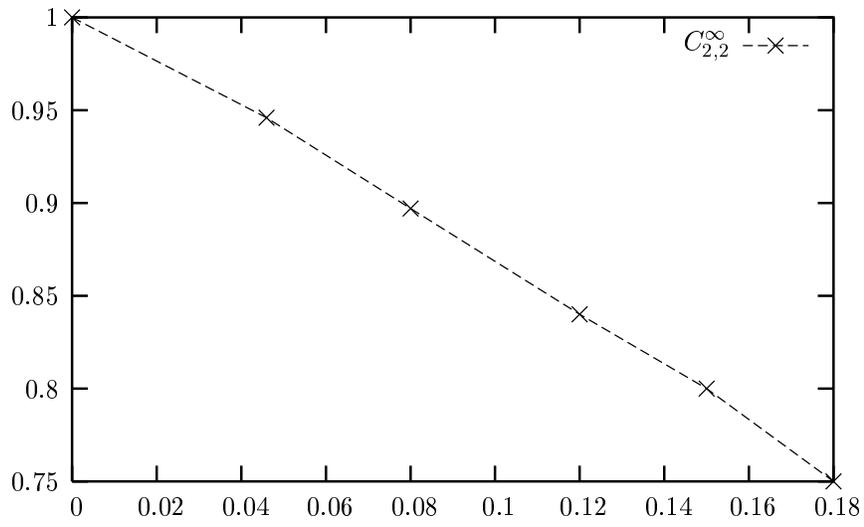,width=\hsize,keepaspectratio,clip}
\caption{Behaviour of $C_{2,2}(\infty)$ vs. $\zeta_R(4)-2\zeta_R(2)$.}
\label{fig6b}
\end{center}
\end{figure}

According to (\ref{cpq}),
the asymptotic value of $C_{p,q}(m)$ for large $m$ is a function of the intermittency, i.e. $C_{p,q}(\infty)$ becomes smaller for larger value of the anomaly
$\zeta_R(p)+\zeta_R(q)-\zeta_R(p+q)$. 
The qualitative prediction of (\ref{cpq}) has been checked against numerical simulations of time dependent random multiplicative
process. In Figure (\ref{fig6b}) we plot the asymptotic 
value of $C_{2,2}(\infty) = 
lim_{m \rightarrow \infty}C_{2,2}(m) $ for a log normal $P(A)$ as 
a function of $\zeta_R(4) -2\zeta_R(2)$. As one can see
the qualitative prediction is confirmed.

\begin{figure}
\begin{center}
\epsfig{file=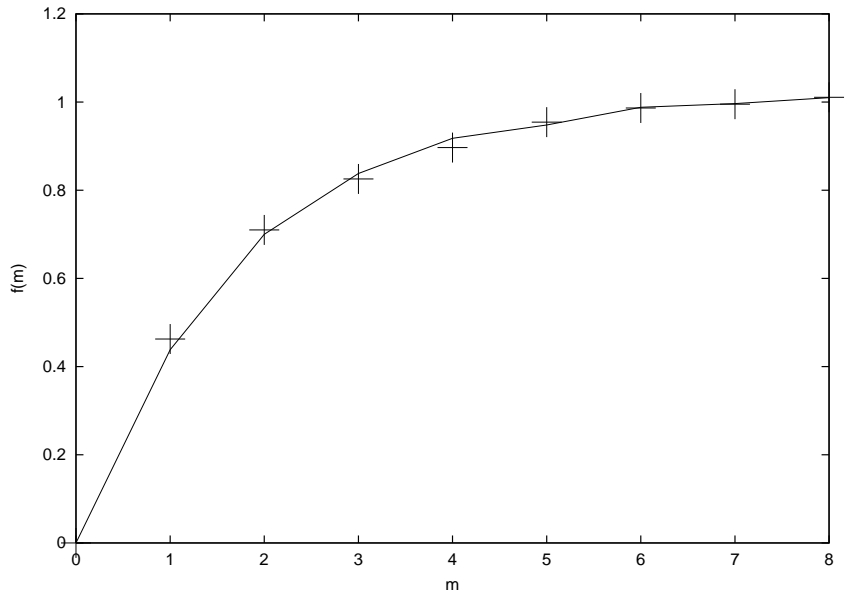,width=\hsize,keepaspectratio,clip}
\caption{Behaviour of $f(m)$ as a function of $m$ for a log-normal distribution and two different values of $\sigma$, 
namely $\sigma = 0.03$ (crosses) and $\sigma = 0.05$ (full line).} 
\label{fig6f(m)}
\end{center}
\end{figure}

A further inspection of the numerical simulations reveals that $C_{p,q}$ can be written as 
\be
C_{p,q}(m) = 1 - \Delta_{p,q} f(m)
\ee
where $f(0) = 0$ and $ f(\infty) = 1 $. While $ \Delta_{p,q} $ is a function of $\zeta_R(p)+\zeta_R(q)-\zeta_R(p+q)$, we
may wonder whether the function $f(m)$ is somehow universal. Although a definite conclusion cannot be reached by looking at the numerical simulations,
still our results seem to indicate that $f(m)$ is either universal or is a function weakly dependent on intermittency. 
This can be seen from Figure (\ref{fig6f(m)})
where we plot $f(m)$ for different  values of $\zeta_R(4)-2\zeta_R(2)$ .

\begin{figure}
\begin{center}
\epsfig{file=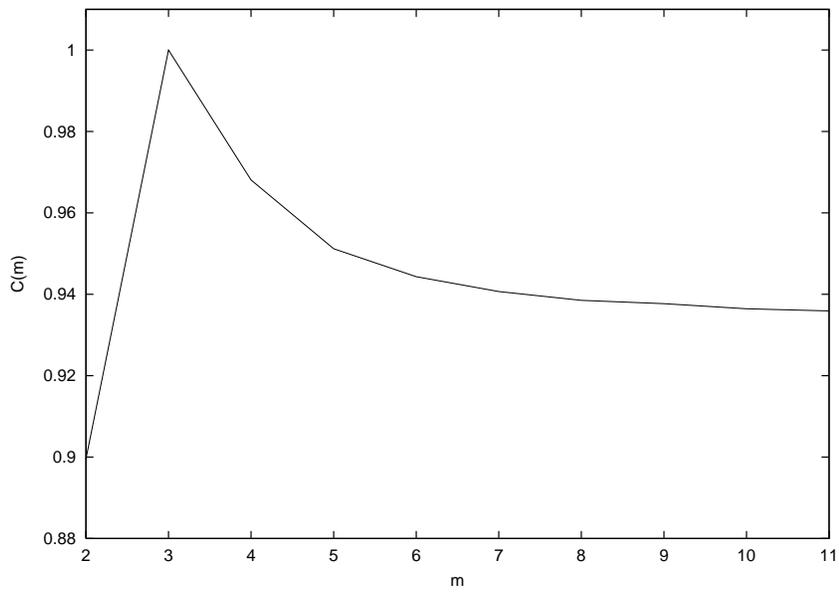,width=\hsize,keepaspectratio,clip}
\caption{Plot of the fusion rule coefficient  $G(m)$ , as defined in the text, for the passive scalar and for $m=-1,..11$.}
\label{passivo2}
\end{center}
\end{figure}

We already observed at the beginning of this Section that a time dependent multiplicative process for a passive scalar does not 
show a re-normalization of the scaling exponents $H(p)$. However, the argument used to derive (\ref{cpq}) can be applied even if the random times
$\tau_n$ are not correlated to the multiplicative process, as in the case of the passive scalar. 
In Figure (\ref{passivo2}),  we show the quantity $G(m)$ defined by the relation: 
\be
\la \theta_3^2 \theta_{3+m}^2 \ra = G(m) \frac{\la \theta_{3+m}^2 \ra}{\la \theta_3^2 \ra} \la \theta_3^2 \ra
\ee
for the passive scalar and $m=-1..11$.
As expected, the fusion rules are satisfied only asymptotically with prefactors $G(m)< 1$.

Let us also  notice  that the coefficients $C_{p,q}$ of the fusion rules are not fully defined, in terms of the time dependent random multiplicative process.
Let us consider the variable 
$$ 
u_n = g_n w_n
$$
where $g_n$ is a random variable, independent of $n$, with the same probability distribution for any $n$,
 i.e.  $\la g_{n_1}g_{n_2}..g_{n_k}\ra=\la g_{n_1}\ra \la g_{n_2}\ra..\la g_{n_k}\ra$ and $P(g_n)$ does not depend on $n$. It follows that
the fusion rules for $u_n$ satisfy:
$$ \la u_{n+m}^pu_n^q\ra = \la g^p\ra \la g^q\ra \la w_{n+m}^pw_n^q\ra $$
which gives
$$
\la u_{n+m}^pu_n^q\ra=  
\la g^p\ra \la g^q\ra C_{p,q} \frac{\la u_{m+n}^p\ra}{\la u_n^p\ra} \la u_n^{p+q}\ra \frac{1}{\la g^{p+q}\ra}
$$
Note that, for each $n$, $w_n$ is the time dependent multiplicative process and that the scaling properties of $u_n$ and $w_n$ are
the same. The above equation implies that the fusion rules coefficients
$C_{p,q}$ for $u_n$ becomes:
\be
C_{p,q} \rightarrow C_{p,q} \frac{\la g^p\ra \la g^q\ra}{\la g^{p+q}\ra} \;\;\; w_n \rightarrow g_n w_n
\label{fusiong}
\ee
Equation (\ref{fusiong}) implies that the asymptotic value of 
$C_{p,q}$ for large $m$ is fixed up to a number (less than $1$) linked to a scale invariant probability distribution function.

Before closing this Section we would like to notice that the renormalized exponents $\zeta_R(p)$ are associated to a less intermittent 
field with respect to the bare exponents $\zeta_0(p)$. This can easily be deduced by using equation (\ref{newzp}). Thus we can predict that
whenever the random times $\tau_n$ are not correlated to $w_n$ we should observe (for a given $P(A)$) an increase of intermittency. 
One may wonder whether this qualitative prediction may have any experimental evidence. This is indeed the case. Let us consider
a shear flow. It has been noted that whenever the mean shear ${\cal S}$ is large enough intermittency increases. For large scale the characteristic time
scale of the dynamics should be dominated by the shear effect  and we would expect the characteristic time scale independent on the scale \cite{toschi1,toschi2,piva}.
In this case, therefore, if we describe intermittency as a random multiplicative process, time dynamics does not lead to a re-normalization
of the scaling exponents, i.e. we should observe the scaling exponents $\zeta_0(p)<\zeta_R(p)$. It is suggestive to think that the
increase of intermittency in shear dominated flows may be understood in terms of the absence of re-normalization in time dependent multiplicative process.

\section{Conclusions and discussions}
We have introduced  a simple multiplicative process 
which embeds intermittency both in time and in space.
This allow us to generate a signal which respect the constraint (imposed by the Navier-Stokes equations) given by equation (\ref{nscale}): $\partial_t w_n \sim l_n^{-1} w_n^2$.
This models is a generalization of the multiplicative process (\ref{rpm}) and
a practical implementation of a signal satisfying the multifractal representation (\ref{time}).
Studying the numerical process we found a ``re-normalization'' of the scaling exponents in space due the non-trivial interplay between multipliers re-generation and time evolution.
We have shown that in the case of a passive scalar this effect is not present.
Furthermore, we have clearly connected the asymptotic gap, observed on fusion rules, with the intermittent scaling exponents.
This finding will be exploited in a forthcoming paper \cite{bbt2}
to build up a stochastic closure to  compute anomalous scaling exponents for  shell models of  turbulence.

We acknowledge G. Boffetta and A. Celani for many discussion in a early stage of this work.

\clearpage
%%%%%%%%%%%%%%%%%%%%%%%%%%%%%%%%%%

%%%%%%%%%%%%%%%%%%%%%%%%
%%
%% Figures
%%
%%%%%%%%%%%%%%%%%%%%%%%%
\clearpage

\clearpage

\clearpage

\clearpage

\clearpage

\clearpage

\clearpage

\clearpage

\clearpage

\clearpage

\clearpage

\clearpage

\end{document}